\documentclass[
    aps,
    prl,
    twocolumn,
    floatfix,nofootinbib,
    superscriptaddress
]{revtex4-2}

\usepackage{graphicx}
\usepackage{subfigure}
\usepackage[utf8]{inputenc}
\usepackage{times}
\usepackage[breaklinks,colorlinks=true]{hyperref}
\hypersetup{linkcolor=red,citecolor=green,urlcolor=blue}
\usepackage{color}
\usepackage{siunitx}
\usepackage{booktabs}
\usepackage[version=4]{mhchem}
\usepackage{ulem}

\usepackage{amssymb} 
\usepackage{amsmath}
\usepackage{amsthm}
\usepackage{braket}
\usepackage{MnSymbol}
\usepackage{stmaryrd}
\usepackage{mathtools}

\newcommand{\pgi}{Peter Gr\"unberg Institut and Institute for Advanced Simulation,
Forschungszentrum J\"ulich and JARA, 52425 J\"ulich, Germany}

\newcommand{\aachen}{Department of Physics, RWTH Aachen University, 52056 Aachen, Germany}

\newcommand{\mainz}{Institute of Physics, Johannes Gutenberg University Mainz, 55099 Mainz, Germany}

\newcommand{\abs}{\mathrm{\abs}}

\usepackage{tikz}
\usepackage{tikz-cd}
\usetikzlibrary{calc}
\usetikzlibrary{decorations.markings}
\usepackage{xcolor}
\usepackage{graphicx}

\makeatletter
\renewcommand\@biblabel[1]{#1.}
\makeatother
\begin{document}

\setcounter{secnumdepth}{2} 

\title{Orbital Rashba effect as a platform for robust orbital photocurrents}

\author{T. Adamantopoulos}
    \thanks{t.adamantopoulos@fz-juelich.de}
    \affiliation{\pgi}
    \affiliation{\aachen}

\author{M. Merte}
    \affiliation{\pgi}
    \affiliation{\aachen}
    \affiliation{\mainz}

\author{D. Go}
    \affiliation{\mainz}
    \affiliation{\pgi}
    
    \author{F. Freimuth}
    \affiliation{\mainz}
    \affiliation{\pgi}
    
\author{S. Bl\"ugel}
    \affiliation{\pgi}
    
\author{Y. Mokrousov}
    \affiliation{\pgi}
    \affiliation{\mainz}

\date{\today}

\begin{abstract}
Orbital current has emerged over the past years as one of the key novel concepts in magnetotransport. Here, we demonstrate that laser pulses can be used to generate large and robust non-relativistic orbital currents in systems where the inversion symmetry is broken by the orbital Rashba effect. By referring to model and first principles tools, we demonstrate that orbital Rashba effect, accompanied by crystal field splitting, can mediate robust orbital photocurrents  without a need for spin-orbit interaction even in metallic systems. We show that such non-relativistic orbital photocurrents are translated into derivative photocurrents of spin when relativistic effects are taken into account. We thus promote orbital photocurrents as a promising platform for optical generation of currents of angular momentum, and discuss their possible applications.

\end{abstract}

\maketitle






\date{\today}


%
%
%
%
%
%
%
%
%
%
%
%
%
%
%


{\it Introduction.} Over the past years, we have been witnessing an ever growing interest in the origins, properties and prospects of currents of orbital angular momentum, which often appear to accompany the currents of spin~\cite{Bernevig_2005, Tanaka_2008, Kontani_2009, Go_2018, Jo_2018}.
Although suppressed in equilibrium, the orbital degree of freedom can be exploited through the flow of orbital momentum in non-equilibrium -- a property which paves the way for moving from conventional spintronics to the upcoming field of orbitronics~\cite{Go_2021b}. Some of the prospects of orbital currents are associated with their often colossal magnitude, and the fact that they emerge even in light materials -- abundant, but utterly useless within the paradigms of spintronics which rely on strong spin-orbit interaction (SOI).
In the field of orbitronics, the orbital Rashba effect (ORE) has recently emerged as a fruitful playground for understanding and studying orbital properties out of equilibrium~\cite{Park_2011, Park_2012, Park_2013, dgo_srep, Go_2021a}. In particular, it has been demonstrated that the formation of chiral orbital textures, arising as a result of broken inversion symmetry at the surfaces and interfaces, results in an orbital version of Rashba-Edelstein effect and generation of colossal orbital currents, 
which can be efficiently used to generate orbital torques on magnetization~\cite{Go_2020a, Go_2020b}. 

On the other hand, we know very little about the properties of orbital currents in the optical domain.
Generally, the physics of optical generation and properties of laser-induced currents appearing at surfaces and interfaces is attracting considerable attention these days, owing in part to the phenomenon of THz radiation generation with so-called spintronics THz emitters~\cite{Seifert_2016, PapaioannouBeigang}. In the context of spinorbitronics inspired optical effects, properties of spin photocurrents start coming under scrutiny as well~\cite{Sherman_2005, Yin_2019, freimuth_2021,*freimuth2017laserinducedarxiv, Xu_2021, Xiao_2021, Fei_2021, Mu_2021, Merte_FGT, Hayami_2022}.
However, besides the studies of inverse Faraday effect in the orbital channel~\cite{Battiato_2014, Berritta_2016, Zhou_2022} and studies on optically-ignited orbital dynamics~\cite{BA_Ivanov_2017, BA_Ivanov_2023, Grzybowski_2023}, the phenomenon of orbital photocurrents is largely unexplored, both from the side of fundamental properties as well as  their microscopic origins. This is even more surprising given that a stark redistribution of  angular momentum, whose orbital component is bound to be significant, is believed to largely mediate for example the process of demagnetization in laser experiments~\cite{Pastor_2023a, Pastor_2023b}. 
Given the prominent role that orbital currents came to occupy in d.c. domain, it is thus pertinent to explore the prospects of orbital photocurrents driven by laser and optical pulses.
Since the concept of the orbital Rashba effect has been  extremely successful in  pushing forward the field of orbitronics, we may ask a question: can the orbital Rashba effect be taken as a unique starting point for exploring the basic physics of orbital currents driven by light? 

In this work we show that the orbital Rashba effect, which can be used to describe the effect of crystal symmetry breaking in wide classes of noncentrosymmetric materials, captures the key features of orbital photocurrents. We use tight-binding and ab-initio methods to demonstrate that the magnitude of the photocurrents of orbital angular momentum at surfaces can be colossal, clarifying the role of crystal field splitting and orbital Rashba strength for their emergence. We uncover the non-relativistic nature of orbital photocurrents, and highlight their  robust nature which results in stable orbital currents even in systems with vanishing SOI strength. We argue that remarkable properties of orbital photocurrents make them an attractive platform for further advances in the realm of orbital optospintronics -- thus laying a unique road to opto-orbitronics.

\begin{figure*}[ht!]
\begin{center}
\rotatebox{0}{\includegraphics [width=0.95\linewidth]{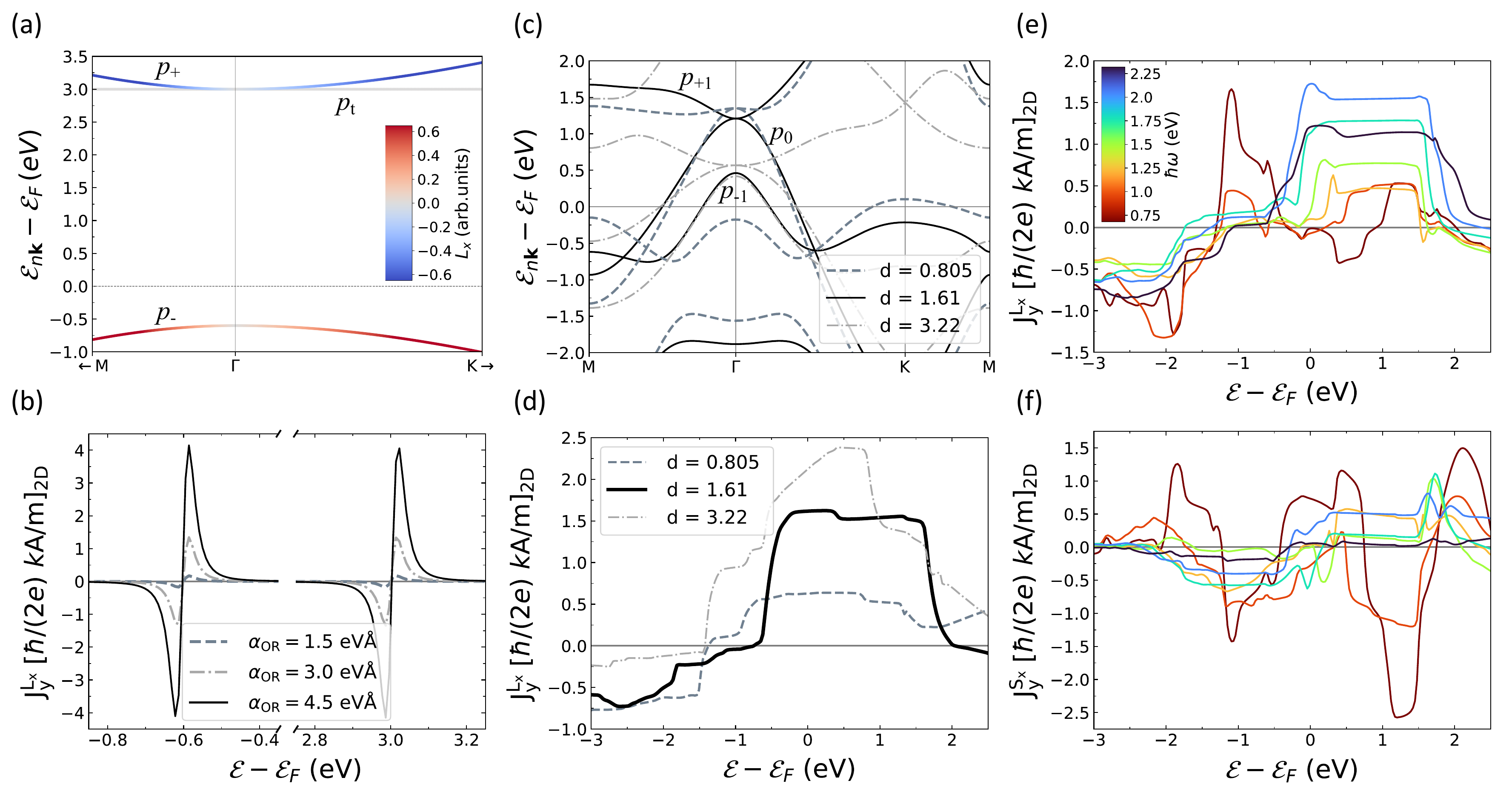}}
\end{center}
\caption{{\bf Orbital photocurrents by orbital Rashba effect}. (a) The orbital polarised band-structure of the $p$-orbitals model. The color of the states marks the value of orbital polarization $L_x$ (color bar). (b) The laser-induced orbital photocurrents within the $p$-orbitals model at $\hbar\omega=3.6$\,eV, in relation to the band filling, for different values of the Rashba constant $\alpha_{\mathrm{OR}}$. (c) Bandstructure of BiAg$_2$ surface without SOI, for different values of separation between Ag and Bi layers (equilibrium value at $d=1.61$\,a.u.). (d) Corresponding dependence of the non-relativistic orbital photocurrent $J_y^{L_x}$ on band filling for $\hbar\omega=2.25$\,eV. 
 (e-f) For the BiAg$_2$ surface with SOI  the orbital (e) and spin (f) photocurrents as shown as a function of band filling, for different frequencies $\hbar\omega$ (color of the line according to the color scale).
 In (b), (d)-(f) the  $J^{L_{x}/S_{x}}_y$ currents arise for light linearly polarised along $x$, propagating along $y$, s with orbital/spin polarisation $x$. 
 }
\label{Fig1}
\end{figure*}

{\it Method.} In this work, to  make qualitative and quantitative predictions, we make use of Keldysh formalism~\cite{freimuth_2016, freimuth_2021,*freimuth2017laserinducedarxiv} to calculate the second order orbital and spin photocurrents which emerge as a response to a continuous laser pulse of frequency $\omega$ according to:
$
    J^{\mathcal{O}_{s}}_{i}=-\frac{a_{0}^{2} I}{\hbar c}\left(\frac{\mathcal{E}_{\mathrm{H}}}{ \hbar \omega}\right)^{2} \operatorname{Im} \sum_{j k} \epsilon_{j} \epsilon_{k}^{*} \varphi^{s}_{i j k}$,
where $a_0$ is the Bohr's radius, $e$ is the elementary charge, $I$ is the intensity of the pulse, $\hbar$ is the reduced Planck constant, $c$ is the light velocity, $\mathcal{E}_H=e^2/(4\pi \epsilon_0 a_0)$ is the Hartree energy, and $\epsilon_j$ is the $j$'th component of the polarization vector of the pulse. The tensor $\varphi^s_{ijk}$ is expressed in terms of the Green functions of the system, two velocity operators ($v_i$ and $v_j$)  and $i$'th component of orbital velocity operator, 
defined as $j^{L_{s}}_{i}=\frac{1}{2}\{v_i,L_s\}$, or the $i$'th component of the spin velocity operator $j^{S_{s}}_{i}=\frac{1}{2}\{v_i,S_s\}=\frac{\hbar}{4}\{v_i,\sigma_s\}$ where the orbital/spin polarization is aligned along direction specified by $s$ via the component of the orbital (spin) angular momentum operator $L_s$ ($S_s$). 
In our formalism, parameter $\Gamma$ is a constant lifetime broadening  introduced so as to  describe the effect of disorder on the electronic states. In this work, we use the values of $\Gamma$ of 25\,meV and intensity of light of 10\,GW/cm$^2$.

{\it Model arguments.} We start by considering a tight-binding model of $p$-orbitals on a two-dimensional triangular lattice (in the $xy$-plane) exposed to a surface potential gradient which mimics the effect of broken inversion symmetry. This model was previously introduced in Ref.~\cite{Petersen_2000} in order to explain the spin Rashba effect at $(111)$-surfaces of noble metals, and it was used recently to study orbital magnetism arising from the orbital Rashba effect~\cite{Park_2012, Kim_2012, Hong_2015}. 
We are interested in the behavior near the $\Gamma$-point for the case when SOI is absent. In this case, the effective Hamiltonian contains diagonal terms which express the energy separation of the $p_x$ and $p_y$ bands from the $p_z$ band due to the crystal field splitting (CFS), while the off-diagonal terms account for the orbital Rashba effect in the general form (an analytic derivation of the tight-binding model is presented in the Supplemental Material)
\begin{equation}
    H_{\mathrm{OR}}(\mathbf{k}) = \frac{\alpha_{\mathrm{OR}}}{\hbar} \mathbf{L}^{(p)} \cdot (\hat{\mathbf{z}} \times \mathbf{k}),
\end{equation}
where $\alpha_{\mathrm{OR}}$ is the orbital Rashba constant, and $\mathbf{L}^{(p)}$ is the vector of matrices of OAM operator in the $p$-basis. The basis set is chosen as $|\phi_{n\mathbf{k}}\rangle = \frac{1}{\sqrt{N}} \sum_{\mathbf{R}} \mathrm{e}^{i\mathbf{k}\cdot\mathbf{R}} |\phi_{n\mathbf{R}}\rangle$, with $\phi_{n\mathbf{R}}$ denoting the $n$-th atomic orbital ($n=p_{x},p_{y},p_{z}$), positioned at the Bravais lattice $\mathbf{R}$, while $N$ numbers the lattice sites. Superpositions of $|\phi_{p_z\mathbf{k}}\rangle$ and of a radial state 	$| \phi_{p_r \mathbf{k}} \rangle = \frac{k_x}{|\mathbf{k}|} | \phi_{p_x \mathbf{k}} \rangle + \frac{k_y}{|\mathbf{k}|} | \phi_{p_y \mathbf{k}} \rangle$ lead to the formation of a chiral OAM texture  whereas a tangential state $| \phi_{p_t \mathbf{k}} \rangle = \frac{k_y}{|\mathbf{k}|} | \phi_{p_x \mathbf{k}} \rangle - \frac{k_x}{|\mathbf{k}|} | \phi_{p_y \mathbf{k}} \rangle$ carries no OAM. 
The $|\phi_{p_\pm \mathbf{k}} \rangle$ states, which arise from the superposition of $| \phi_{p_z \mathbf{k}} \rangle$ with the radial state, carry chiral OAM which is evaluated analytically by
\begin{equation}\label{eq:oam}
    \langle \phi_{p_\pm \mathbf{k}} | \mathbf{L}^{(p)} | \phi_{p_\pm \mathbf{k}} \rangle = \pm \hbar \frac{2 \alpha_{\mathrm{OR}}}{\Delta_{\mathrm{CF}}} \hat{\mathbf{z}} \times \mathbf{k},
\end{equation}
where $\Delta_{\mathrm{CF}}$ is the value of CFS. The chiral distribution of OAM in $k$-space is the trademark of ORE. A typical bandstructure of the model is shown in Fig.~\ref{Fig1}(a) in the vicinity of the $\Gamma$-point~[Suppl].
The distance between the $p_\pm$ states is tuned by the values of $\alpha_{\mathrm{OR}}$ and $\Delta_{\mathrm{CF}}$, whereas $\alpha_{\mathrm{OR}}$ also influences the curvature of the $p_\pm$ bands. As expected from Eq.(\ref{eq:oam}), the $p_\pm$ states have OAM of opposite sign,  
while the $p_t$ state  carries no OAM.

It is known that the ORE unleashes the linear in the field current-induced orbital magnetization~\cite{Go_2021a}. In the next step, we  utilize the Keldysh formalism to demonstrate the generation of laser-induced orbital photocurrents within the ORE. In Fig.~\ref{Fig1}(b) we show the only symmetry-allowed component of the orbital photocurrent $J^{L_x}_y$ as a function of band filling.
The light frequency is set at $\hbar\omega=3.6$\,eV and matches exactly the energy difference between the $p_\pm$ states at $\Gamma$. The calculated orbital response is remarkably sharp and strong with the peaks located slightly above and below 
the energies of the $p_\pm$ states. 
Since the ORE drives orbital photocurrents already without SOI, the effect should strongly depend on the strength of the orbital Rashba effect. 
Indeed, as shown in Fig.~\ref{Fig1}(b), the orbital photoresponse increases drastically with the value of  $\alpha_{\mathrm{OR}}$. 
For the value of $\alpha_{\mathrm{OR}}=3.0$\,eV\AA\, which corresponds to the case of BiAg$_{2}(111)$ surface alloy~\cite{BiAg2(111)_surf_alloys}, the predicted magnitude of the orbital photocurrent reaches a very large value of about 1600\,$\frac{\hbar}{2e}\frac{\rm A}{\rm m}$. To check whether such values are  achievable in 
realistic materials, we next turn to first principles calculations. 

{\it Realistic systems.} To explore the possibility of generating large orbital photocurrents in realistic systems, we choose a prominent representative of the class of strong Rashba surfaces: the BiAg$_2$ alloy~\cite{BiAg2(111)_surf_alloys, Carbone_2016}. We compute the
electronic structure of the BiAg$_2$ surface from ab-initio by using the the full-potential linearized augmented plane wave \texttt{FLEUR} code~\cite{fleur}, and
employ the Wannier interpolation technique
to assess from first principles the laser-induced orbital and spin photocurrents by using Keldysh formalism
~\cite{Pizzi2020,Merte_FGT, Adamantopoulos_2022}. Further computational details and details of the method are given in the Supplemental Material. The symmetry behavior of the calculated orbital photocurrents agrees with the symmetry predictions of the effective non-magnetic spin Rashba model~\cite{freimuth_2021,*freimuth2017laserinducedarxiv}. For simplicity, in the following we restrict our discussion only to  $J^{L_{x}}_{y}$ and $J^{S_{x}}_{y}$ components of the orbital and spin photocurrents, respectively, which arise for light linearly polarised along the $x$-axis.



We begin our discussion by studying the case without SOI.
For the BiAg$_2$ surface it is known that the $sp$-orbital hybridization gives rise to the prominent ORE, which is reflected in its characteristic bandstructure, shown with black lines in Fig.~\ref{Fig1}(c) for the equilibrium lattice constant: the main features are identical to that of the orbital Rashba model, with $p_{+1}$ and $p_{0}$ bands of Bi  degenerate at $\Gamma$-point and split off from the $p_{-1}$ band. Remarkably, we observe that, in accordance to the predictions of the model analysis from above, the ORE gives rise to very large {\it non-relativistic} orbital currents in response to laser light, as shown for the case of $\hbar\omega=2.25$\,eV as a function of band filling in Fig.~\ref{Fig1}(d). Here, for the equilibrium structural configuration, we not only observe a magnitude of the order of 1500\,$\frac{\hbar}{2e}\frac{\rm A}{\rm m}$ for $J_y^{L_x}$, but also witness a remarkable robustness of the orbital current to the changes in band filling, reminiscent of a band insulator, where such behavior is expected when the Fermi energy is varied within the band gap~\cite{Canonico_2020, Cysne_2021, Cysne_2022, Zeer_2022}.  

Before proceeding with the detailed $k$-resolved analysis of the origin of the giant orbital photocurrents and their robust behavior, we note that the width of the observed plateau in energy can be roughly associated with the difference in energy between $p_{-1}$, $p_{0}$ and $p_{+1}$ bands: the plateau starts from the region of flat $p_{-1}$ and $p_{0}$ states around $[-1;-0.5]$\,eV and goes up to the energy region of flat $p_{+1}$ states around $+1.6$\,eV. Outside of this region, the orbital response is generally suppressed.
We demonstrate that the $p_{\pm 1}$ separation is correlated with the orbital current plateau by changing the distance between the Ag and Bi layers, which results in modifications in the effective orbital Rashba strength and crystal field splitting, in turn having a strong effect on the energy dependence of $J_y^{L_x}$. Indeed, as we can see in Fig.~\ref{Fig1}(d), the size of the plateau follows the effective separation between the orbitally-polarized bands: as the distance is increased by a factor of two, the $p_{+1}$-band goes down in energy with respect to $p_{-1,0}$ states, but not as significantly as for the case of decreased inter-layer distance. In both cases, while the changes in the magnitude of the orbital currents are driven by modifications in the orbital Rashba constant, as predicted by the model, the modifications in the crystal field splitting together with $\alpha_{\rm OR}$ reduce the effective size of the plateau.

\begin{figure*}[t!]
\begin{center}
\rotatebox{0}{\includegraphics [width=0.92\linewidth]{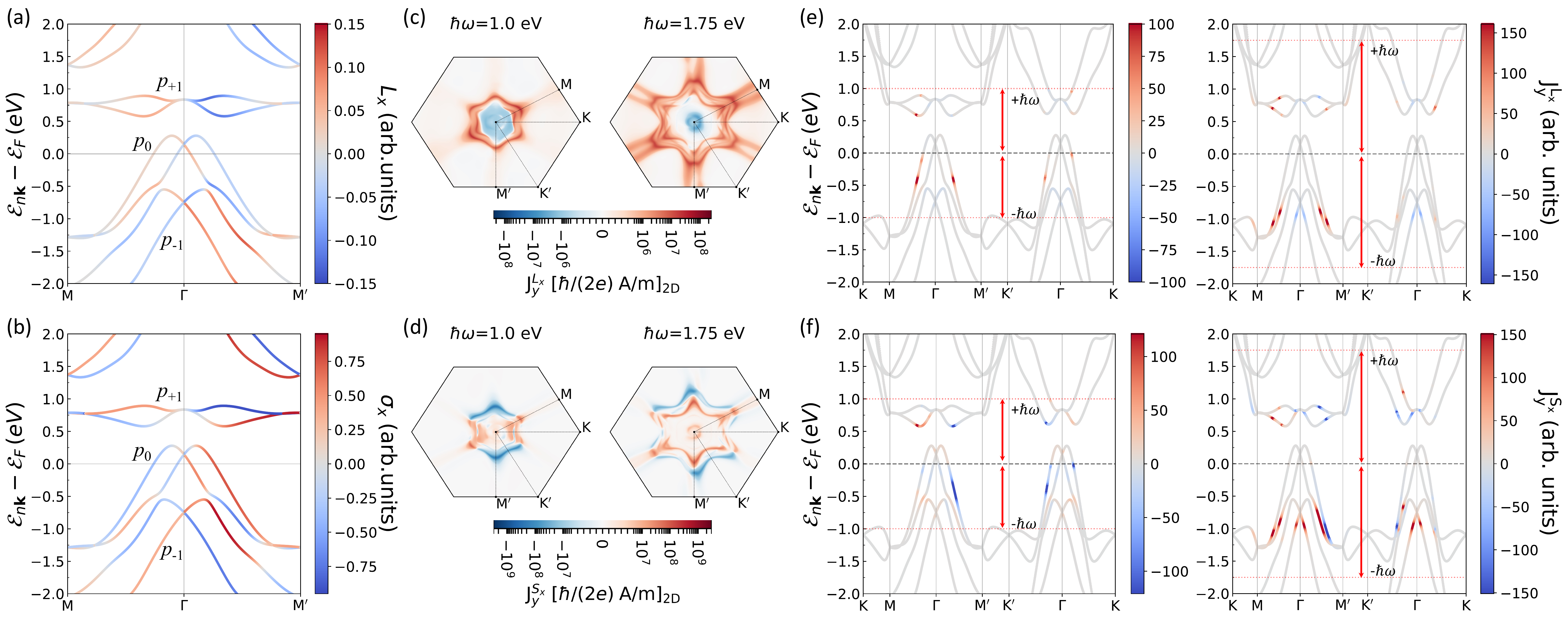}}
\end{center}
\caption{{\bf Microscopics of orbital photocurrents}. (a), (b) Orbital (a) and spin (b) polarised band-structure of the BiAg$_2$ surface with SOI, with the states of distinct orbital symmetry marked as $p_{\pm1}$, $p_0$. (c), (d) Reciprocal space distribution of integrated  (c) orbital and (d) spin photocurrents. (e), (f) Band-resolved two-band contributions of (e) orbital and (f) spin photocurrents. In all cases the responses are calculated for two different light frequencies of $\hbar\omega=1.00$\,eV (left panels in (c)-(f)) and $\hbar\omega=1.75$\,eV (right panels in (c)-(f)), over the constant energy surface of the shifted Fermi level $\mathcal{E}_{F}^{\prime}=\mathcal{E}_{F}+0.9$\,eV. The presented component arises for light linearly polarised along the $x$-axis and flows along the $y$-axis with orbital/spin polarisation along the $x$-axis. In (e), (f) the horizontal dotted red lines at $\pm1.0$\,eV (left panels) and at $\pm1.75$\,eV (right panels) denote the energy of the laser pulse.
}
\label{Fig2}
\end{figure*}


While it is clear that we can achieve gigantic non-relativistic orbital currents, it is important to explore the role that the spin-orbit interaction (SOI), universally present in real solids, can play. In particular, we want to answer the following questions, which are particularly relevant for our heavy test system: (i) What is the impact of SOI on magnitude and robustness of the orbital photocurrents? (ii) Would the orbital currents become irrelevant when SOI-driven spin photocurrents come into play? In particular the latter have been recently shown to become very prominent in two-dimensional materials with broken inversion symmetry~\cite{Adamantopoulos_2022, Merte_FGT, Fei_2021, Xu_2021, Xiao_2021, Mu_2021}.  


To answer the first question, we take into account SOI and compute the orbital photocurrents for various frequencies, presenting the results as a function of band filling in Fig.~\ref{Fig1}(e). By comparing the $J_y^{L_x}$ curves obtained for $\hbar\omega=2.25$\,eV with and without SOI, we remarkably find little difference. This signifies that orbital currents are much less sensitive to the changes in the structure of the bands brought by SOI, than to much more decisive effect of CFS and ORE, which determine the main features of the bands on a larger than SOI energy scale. Upon decreasing the frequency, not all transitions across differently-polarized $p$-states can be activated, and thus the width of the plateau becomes smaller while the finer features of the $p$-states come into play. For example, a larger peak in $J_y^{L_y}$ just above $-1$\,eV at $\hbar\omega=0.75$\,eV can be explained by transitions among $p_0$ and $p_{-1}$ states around that energy which are not active at higher frequencies.  

A more drastic effect of SOI lies in promoting the spin physics within the Rashba realm: if without SOI each state of the system is spin-degenerate, SOI splits this degeneracy which makes the overall electronic structure spin-polarized in $k$-space $-$ a picture often referred to as the (spin) Rashba effect~\cite{spin_Rashba}. As a result, the combined effect of ORE and SOI unleashes the photocurrents of spin arising in response to light~\cite{freimuth_2021,*freimuth2017laserinducedarxiv}. To compare the orbital photocurrents to photocurrents of spin in BiAg$_2$, we compute the latter, presenting the results in Fig.~\ref{Fig1}(f). 
From comparison of the two types of currents we conclude that although overall the magnitude of the spin and orbital photocurrents is nominally similar, in the robust region of band filling between 0\,eV and +2\,eV the orbital currents are dominating by far, except for smaller frequencies of light. For example, for $\hbar\omega=2.0$\,eV the achieved orbital photoconductivity value of $\sigma^{yL_{x}}_{xx}\approx3000$\,($\hbar$/2e)\,${\mu}$A/V$^2$ can be compared to a corresponding spin photoconductivity magnitude of $\sigma^{yS_{x}}_{xx}\approx500$\,($\hbar$/2e)\,${\mu}$A/V$^2$. One has to remark here that, upon reducing the SOI strength by hand in the calculations, the magnitude of the spin currents is respectively reduced (not shown), vanishing in the limit of zero SOI. Given that among Rashba materials BiAg$_2$ is among the heaviest, at lighter Rashba surfaces, such as that of oxygenated Cu~\cite{Go_2021a}, the orbital character will dominate the nature of the photoinduced currents of angular momentum.

At this point we are ready to get a deeper insight into the properties of orbital photocurrents by scrutinizing  their anatomy in reciprocal space. We start by looking at the distribution of orbital and spin photocurrents in $k$-space, assuming the shifted Fermi energy of 0.9\,eV above the true Fermi level -- which falls right into the middle of the plateau region and roughly corresponds to the position of $p_{+1}$ band -- and vary the frequency from 1.0 to 1.75\,eV, Fig.~\ref{Fig2}(c,d).
In Fig.~\ref{Fig2}(c) and Fig.~\ref{Fig2}(d) we present the reciprocal space distribution of the orbital and spin photocurrents, $J_y^{L_x}$ and $J_y^{S_x}$,  respectively. At first glance, the distributions look quite similar in structure: upon increasing the frequency, the regions of strong contributions move further away from the $\Gamma$ point, which is a signature of active electronic transitions moving closer to the Brillouin zone boundaries and high-symmetry lines. Clearly, here the distribution of the  photocurrents of angular momentum probes the shape of the valence band which launches the transitions into the $p_{+1}$ band. One remarkable feature sticks out, however: while at larger frequencies the sign of the orbital photocurrents is predominantly uniform throughout the Brillouin zone, the photocurrents of spin exhibit a characteristic behavior where the regions of positive and negative contributions reflect each other in shape. Such an ``onion skin" effect explains why, in contrast to orbital photocurrents, the spin photocurrents are strongly suppressed when integrated over the $k$-space. 

The origin of the key difference between two types of currents becomes clear after analyzing in detail the electronic structure.  In  Fig.~\ref{Fig2}(a) and Fig.~\ref{Fig2}(b) we show the bands of BiAg$_2$ with SOI along M$\Gamma$M$^\prime$, colored by the magnitude of the orbital (a) and spin (b) polarization of the states along $x$. In comparison to non-relativistic bandstructure, presented in Fig.~\ref{Fig1}(c), we clearly see a very prominent splitting of the states arising as a result of very large SOI of the system. Remarkably, we observe that while the SOI splitting keeps the sign of the orbital polarization of the SOI-split states the same along a given $k$-path, on the contrary, the spin-polarization of the states alternates in sign. This effect, particularly clear for $p_{+1}$ and $p_{-1}$ bands around $+0.75$\,eV and $-1$\,eV respectively, stems from the fact that basic features of the orbital polarization are already set by the combined effect of ORE and CFS onto which SOI is applied as a perturbation. For the case of spin the situation is drastically different: an alternating spin polarization is a direct consequence of the $k$-dependent SOI-mediated coupling of electron's spin to the surface gradient, which is the essence of the spin Rashba effect. As a result, the origin of the orbital photocurrents in ORE enhances their magnitude, while photocurrents of spin are suppressed by the spin Rashba effect -- with suppression becoming stronger as the splitting between the alternating-in-sign regions in $k$-space becoming smaller with decreasing SOI.

The fingerprints of this behavior can be also clearly seen in the band-resolved decomposition of the photocurrents. In Fig.~\ref{Fig2}(e,f) we plot the contributions of so-called resonant two-state transitions~\cite{Zhang_2018, Zhang_2019} to orbital (e) and spin (f) photocurrents along the high symmetry lines: irrespective of light frequency chosen (1.0\,eV or 1.75\,eV), an alternating-in-sign behavior is much more characteristic of a photocurrent of spin than of its orbital counterpart, in accord to the qualitative behavior of state polarization. The latter plots also provide a key to understanding the quantization-like features in  orbital photocurrents, discussed above.  Important thing to realize is that the narrow $p_{+1}$-group of states stands energetically well-separated from the rest of orbitally-polarized valence and conduction bands, which is primarily the consequence of the crystal field splitting at the ORE surface. The role of light with a given frequency is then to launch electronic transitions among the $p_{+1}$ band and occupied states which are lower by about $\hbar\omega$ in energy, see Fig.~\ref{Fig2}(e,f). Respectively, changes in frequency probe the structure of the valence $p_0$- and $p_{-1}$-states -- an effect we speculated to take place above. On the other hand, since the energetics of electronic transitions is already set by the magnitude of CFS and light frequency, the variation of band filling within a certain interval plays only a minor role, which gives rise to a robust plateau in the orbital photocurrent.
Note that  quasi-quantization plateaus in orbital  photocurrent have  little to do with the presence of a global band gap in the spectrum apparent in Fig.~\ref{Fig2}, since the robust behavior emerges already in non-relativistic limit where the system is manifestly metallic.


{\it Discussion.} Given the observed properties of orbital photocurrents, it is tempting to imagine what impact they could have in optical magnetism. First of all, the role of orbital photocurrents for demagnetization properties has to be explored. In fact, first experiments seem to suggest that an interaction of magnetization with laser pulses may give rise to orbital currents~\cite{Fert_2022} which can even exhibit a long-range nature~\cite{Seifert_2023}. We thus speculate that orbital photocurrents may interact with magnetic properties in an unexpected way. For example, would it be feasible to exploit orbital photocurrents for exerting sizeable optical torques on the magnetization which take place on the femtosecond timescale? This may prove indispensable in~e.g.~all-optical antiferromagnetic switching schemes which do not rely  on intrinsically slower demagnetization processes. The non-relativistic nature of orbital photocurrents would, on the other hand, open a whole new palette of opportunities associated with material realization of optical phenomena, traditionally escaping the search as generators of sizeable spin-based processes. At the same time, the origin of orbital photocurrents in the  crystal structure and key features of chemical composition would allow for a reliable and robust control of opto-orbital properties, promoting new possibilities in our struggle to master  light-matter interaction. 

{\it Acknowledgements.} This work was supported by the Deutsche Forschungsgemeinschaft (DFG, German Research Foundation) $-$ TRR 173/2 $-$ 268565370 (project A11), TRR 288 – 422213477 (project B06), and the Sino-German research project DISTOMAT (MO 1731/10-1). This project has received funding from the European Union’s Horizon 2020 research and innovation programme under the Marie Skłodowska-Curie grant agreement No 861300.
We  also gratefully acknowledge the J\"ulich Supercomputing Centre and RWTH Aachen University for providing computational resources under projects  jiff40 and jara0062.


\hbadness=99999 
\bibliography{literature}




\end{document}